# The benefits of completing homework for students with different aptitudes in an introductory physics course

F. J. Kontur and N. B. Terry

*Department of Physics, United States Air Force Academy, USAF Academy, Colorado 80840, USA*

We examine the relationship between homework completion and exam performance for students having different physics aptitudes for five different semesters of an introductory electricity and magnetism course. In our analysis, we plot exam scores versus homework completion scores and calculate the slopes of the line fits and the Pearson correlations. On average, completing many homework problems correlated to better exam scores only for students with high physics aptitude. Low aptitude physics students had a negative correlation between exam performance and completing homework; the more homework problems they did, the worse their performance was on exams. One explanation for this effect is that the assigned homework problems placed an excessive cognitive load on low aptitude students. As a result, no learning or even negative learning might have taken place when low aptitude students attempted to do assigned homework. Another explanation is based on the fact that the negative benefit effects first appeared when magnetism concepts were introduced. According to this explanation, low aptitude students had difficulty consolidating knowledge of magnetic fields with previously-learned knowledge of electric fields. A third possibility, that a high homework copying rate by low aptitude students impeded learning, is rejected because two different analyses revealed no evidence of homework copying.

## I. INTRODUCTION

Homework is a key part of nearly every college-level physics course. Therefore, it is not surprising that homework has been one of the most well-studied aspects of physics pedagogy. Numerous articles have examined the advantages and disadvantages of online homework [see, for example, 1-5]; other topics include the deficiency of traditional homework in teaching physics concepts [6] and ways to deal with homework copying [7-9]. Despite the large amount of research that has been done on homework in physics classes, there are still many questions that remain about how it can best be used to aid in student learning. The present study examines one of those remaining questions – what are the benefits of completing homework for introductory physics students with different aptitudes? While at least two articles [10-11] touch on this subject, to our knowledge, this is the first study which directly examines the question. As societal needs and public policy push a larger number of students into pursuing science and technology careers, it will become increasingly important for physics educators to know how beneficial their pedagogical tools are for students having a wide range of abilities. We hope that this investigation will be helpful for teachers dealing with such issues in their classrooms.

Previous studies have examined the benefits of doing homework for students of different backgrounds and skill levels in introductory physics courses. Cheng *et al.* looked at the differences in student learning when two different types of homework (ungraded homework and graded online homework) were combined with two different types of teaching methods (interactive and non-interactive) in different sections of introductory physics [10]. One way that they categorized their results was by students' final grades in the course. For the non-interactive sections, they found that students who received an A or B in the course had higher gains on the Force Concept Inventory (FCI) [12] when they had online graded homework compared to A and B students who had ungraded homework. However, the FCI gains of C and D students in the non-interactive sections showed no statistical difference based on the type of homework administration used. Conversely, for interactive sections, all students had higher FCI gains when they used the online graded homework compared to similar students who had ungraded homework. Morote and Pritchard did a related study for introductory physics courses in which they measured the correlation between various course assessments and 12 different variables related to student background [11]. They found statistically significant correlations between test scores and the previous level of calculus that students had taken. In addition, students' final exam scores had a statistically significant correlation to the previous level of physics that they had taken. However, they found that performance on a web-based homework tutorial, myCyberTutor, showed no statistically significant correlation with any student background variables. This implied that the myCyberTutor homework was, in a sense, a less biased measurement of student performance in the physics course than exam scores, which depended on student background.

The studies discussed in the previous paragraph offer suggestions for how homework may benefit students with different levels of physics aptitude, but they lack in-depth analysis of the connection between homework completion and learning in physics courses. For example, Cheng *et al.* find that interactive classes combined with graded online homework led to the highest learning gains for all students. However, their data do not indicate what aspects of online homework and interactive classes led to increased learning gains. In their work, Morote and Pritchard find that students have uniform success at completing myCyberTutor homework regardless of academic background. However, there is not a similar uniformity of success on exams. It is not clear why, for students having weaker academic backgrounds, performance on homework does not translate into the same level of performance on exams compared to their classmates with stronger academic backgrounds.

The present study examines the link between homework completion and exam performance for students with different physics aptitudes in an attempt to answer these questions. For the spring 2009 – spring 2011 introductory electricity and magnetism courses at the United States Air Force Academy (USAFA), we bin students by incoming aptitude based on their grades in pre-requisite courses. We then examine how much students at each aptitude level benefit from completing online homework, where benefit is defined by their performance on course exams that have a combination of traditional workout problems and conceptual multiple-choice questions. Our findings indicate that, while students with high incoming physics aptitude get some benefit on course exams from completing homework, students with medium incoming physics aptitude get no benefit on course exams from completing homework. Even more astounding, we find that students with low incoming physics aptitude perform more poorly on exams when they complete a greater number of homework problems than their peers in the same aptitude group. We consider several explanations for these observations. One explanation is that homework imposed an excessive cognitive load on low aptitude students. The other explanation is that learning magnetism concepts negatively interfered with previously-studied knowledge of electricity concepts for low aptitude students. We find that homework copying does not seem to be a likely explanation.

## II. COURSE OVERVIEW

All USAFA students are required to take two semesters of calculus-based introductory physics in order to graduate. The first-semester course covers Newtonian mechanics concepts. The second-semester course covers electricity and magnetism concepts as well as basic optics, and will hereafter be referred to as E&M. The E&M course was chosen for this study for several reasons. The first reason, discussed in Section III, is that we were able to accurately characterize the incoming physics aptitude of students taking E&M based on their grades in previous calculus and physics courses. The second reason is that students are less likely to have had a high school physics course that covers E&M topics than a high school course that covers mechanics topics. Therefore, while there may be a large percentage of students who can be successful in the mechanics course due to their high-school physics background, most of the student learning that is necessary for success in E&M must take place in the USAFA course. The final reason is that the number and complexity of concepts covered in the E&M course is greater than in the mechanics course, which could lead to more divergence between the performances of high aptitude and low aptitude students compared to the mechanics course.

### A. Student population and demographics

During an academic year, approximately 1,000 students take the USAFA introductory E&M course. As mentioned before, all USAFA students must take the two-semester introductory physics sequence in order to graduate, but many of the students who take these courses will go on to major in subjects outside of science and engineering fields. Based on data for the Class of 2013, 46% of USAFA students major in the humanities and social sciences, 44% major in mathematics, the physical sciences, or an engineering field, and 9% major in biology or geospatial science. The students accepted to USAFA typically have strong academic backgrounds. For the Class of 2013, 82% come from the top quarter of their high school class, 53% come from the top tenth of their high school class, and 9% were either the salutatorian or valedictorian of their high school class. Their average math SAT score is 664, their average math ACT score is 30.3, and their average science reasoning ACT score is 29.4. The male:female ratio for the Class of 2013 is 4:1. Students admitted to USAFA must fall in the 17-22 age range. All students live in dormitories on campus, and class attendance is mandatory.

The standard course sequence for USAFA students places mechanics in the spring of their freshman year and E&M in the fall of their sophomore year. Because of this sequencing, the fall semester is the large-enrollment offering of E&M, with between 700-900 students enrolled in the course. The spring semester of E&M is the smaller-enrollment offering, with between 200-300 students enrolled during that semester. There are a number of reasons why students take E&M during the spring semester, but the most common reason is that they have had difficulties in previous math and/or science courses. This is demonstrated by comparisons of the grade point averages (GPAs) in pre-requisite courses for spring versus fall semesters. Spring semester students had an average Calculus I grade of 2.37 (on a 4.0 scale) while fall semester students had an average grade of 2.97. In Calculus II, spring semester students had an average grade of 2.33 while fall semester students had an average grade of 2.74. In mechanics, spring semester students had an average grade of 2.35 while fall semester students had an average grade of 2.60.

Page **3** of 16

TABLE I. E&M topics covered in each class block for the USAFA E&M course.

| Class Block | Corresponding Lessons | Topics Covered |
|---|---|---|
| 1 | 1 - 10 | Coulomb's Law, Superposition, Calculating Electric Fields of Continuous Charge Distributions, Gauss's Law |
| 2 | 11 - 19 | Electric Potential and Energy, Capacitance, Resistance, Kirchoff's Laws |
| 3 | 20 - 30 | Magnetic Forces, Biot-Savart Law, Ampère's Law, Faraday's Law for Induced EMFs and Induced Electric Fields, Maxwell's Addition to Ampère's Law |
| 4 | 31 - 40 | Ray Optics, Thin Lens Equation, Interference, Diffraction, Optical Resolution |

### B. Course structure and content

There are 40 lessons in the E&M course, and the semester is divided into 4 blocks of approximately 10 lessons each, as shown in Table I. Each block focuses on different aspects of E&M. After each of the first three blocks, students take a midterm exam on material from that block. An exception to this was the spring 2009 semester, where students took a midterm after the second and third blocks, but not after the first block. After the fourth block, students take a cumulative final exam.

### C. Pedagogy

The E&M course is taught in sections where the enrollment is set at approximately 20 students. There are, on average, 12 different instructors for the large-enrollment offering of the course and 5 different instructors for the smaller-enrollment offering. Regardless of instructor, all students in the course have the same textbook, use the same syllabus, complete the same assignments, and take the same quizzes and exams. During the semesters considered in this study, the E&M course textbook was the first edition of *Essential University Physics* by Richard Wolfson [13]. The learning objectives for each lesson were selected from the learning objectives given at the beginning of the textbook chapters. Students received completion points for doing pre-class work before every lesson. Pre-class questions focused on example problems and important concepts from the reading. USAFA physics instructors are given extensive training in a variety of interactive teaching techniques, including just-in-time teaching [14], peer instruction [15], think-pair-share [16], and board work problem-solving. This training is done as part of the instructors' new faculty orientation. Instructors are highly encouraged to use these interactive teaching techniques in the E&M course.

### D. Homework

Homework was administered through the Mastering Physics online system. The assigned problems were taken from end-of-chapter problems in the textbook. Students were generally assigned 2-3 homework problems for each lesson, with a total of 90-100 problems being assigned over the duration of the semester. Students were given up to five tries to get the correct answer on homework problems, with no deduction for an incorrect answer until the final attempt. The amount of credit given for homework varied from a low of 6.0% to a high of 9.6% of the total course grade. The Mastering Physics website assigns a difficulty rating for homework problems on a 1-5 scale, with 1 being easiest and 5 being hardest. The problems assigned in the E&M course were equal to or slightly lower in difficulty than the average difficulty of all the end-of-chapter problems for the chapters that were covered.

### E. Examinations

There were three midterms and a cumulative final exam for all but one of the semesters of E&M being studied. The midterms were each worth ~10% of the course grade and the final exam was worth 25%. In the spring 2009 E&M course, there were only two midterms, which were each worth 13.5% of the course grade; the final exam was worth 25%. The midterms are 80- to 110-minute in-class exams which have ten conceptual multiple-choice questions worth 50% of the exam points and two or three homework-type workout problems worth 40-50% of the exam points (7 of the 14 midterms also had a short-answer question that was worth 10% of the exam points). The final exam consists of 30-35 conceptual multiple-choice questions, worth 60-70% of the exam points, and two or three workout problems, worth 23-36% of the exam points. A multiple-part short-answer question was included on the fall 2009 and spring 2011 final exams. On the fall 2009 final exam, the short answer question was worth 6% of the final exam points, and on the spring 2011 final exam, the short-answer question was worth 17% of the exam points.

The first drafts of the exams were written by personnel at USAFA's Center for Physics Education Research. These personnel were not instructors in the E&M course, and they wrote the exams using the course learning objectives. The exams were edited based on feedback from instructors. In general, the editing process involved re-wording of questions to make them more understandable and replacement of questions based on which concepts were emphasized or not emphasized in each block. The overall goal throughout the exam writing and editing process was to test physics understanding rather than memorization or pattern-matching.

The workout and short-answer questions on the exams were graded collectively by all of the course instructors. The instructors were organized into grading groups; each group graded a specific workout problem or short-answer question. The groups started the grading session by developing and calibrating a grading rubric for the problem. The conceptual multiple-choice questions were graded by computer.

### III. METHODOLOGY

In order to determine the learning benefits from completing homework for students with different physics aptitudes, we require a working definition of at least three different terms – student learning, benefit from completing homework, and physics aptitude. The following subsections describe our rationale for how we define each of these items.

### A. Student learning

In physics education research, student learning is often measured using standardized tests, with the most common being the FCI [12] for introductory mechanics courses and the Conceptual Survey of Electricity and Magnetism (CSEM) [17] or the Brief Electricity and Magnetism Assessment [18] for introductory E&M courses. In this study, we chose to use course exams rather than one of these standardized assessments as a way to characterize student learning. We had two main reasons for making this choice. The first reason is that light and optics, which comprises a quarter of the content in our E&M course, is not covered on the standardized assessments. The second reason is that the standardized exams test conceptual knowledge rather than problem-solving ability. While it has been shown that increased conceptual knowledge leads to better problem-solving [15], we wanted to separately identify if homework was benefitting students' conceptual and/or problem-solving abilities.

### B. Student benefit from completing homework

We claim that certain groups of students can benefit from doing homework and that other groups of students show no benefit or negative benefits from doing homework. A group of students is said to benefit from doing homework if members of that group who do more homework than other members of the group tend to have higher exam scores. For this study, we group students by physics aptitude; for each aptitude group we do a line fit for a plot of exam scores versus homework completion scores and determine the slope of the line fit and the Pearson correlation for each plot. The slope of the line fit indicates how much better, on average, students in a given aptitude group would have done on exams it they increased their homework completion by some amount, and the Pearson correlation allows us to determine a *p*-value, which indicates the statistical significance of the data.

### C. Physics aptitude

Because student performance on course exams is our measurement of learning, we determined physics aptitude by finding the variable that, at the beginning of the semester of the E&M course, best predicts student success on E&M exams. There were a number of different variables that we considered – SAT math scores, ACT math and science scores, FCI pre- and post-test scores, CSEM pre-test scores, and grades in the three pre-requisite courses (Calculus 1 and 2 and Mechanics). Table II has a summary of the average Pearson correlations between each of these variables and students' combined scores on the midterms and final exam for E&M. As Table II shows, grades in pre-requisite courses have the largest correlations. The high correlation between calculus grades and E&M exam performance is consistent with previous studies that found a high correlation between students' math skills and their exam grades in college physics [19]. The bottom column of Table II shows that combining the grades in the three pre-requisite courses results in a larger correlation than that obtained by using the grades in any of the individual courses. Therefore, student aptitude was measured using students' combined grade point average in Calculus 1, Calculus 2, and Introductory Mechanics. The highest grade point average that can be achieved in those courses is 4.0, for a student who received an A in all three courses. The lowest grade point average is 1.0, for a student who received a D in all three courses. A student who fails one of the courses must retake it before taking E&M. For simplicity, in the case of students who take a course more than once, we only considered their most recent grade in the course. As shown in Table III, we grouped students into four physics aptitude groups based on their grade point averages in the three pre-requisite courses. Each aptitude group covers 0.75 grade points, from the maximum of 4 to the minimum of 1. For the remainder of this article, the aptitude groups will be referred to using their names in Table III.

TABLE II. Average Pearson correlations between different variables and students' combined scores on the midterm exams and final exam in E&M. The average is taken over the five semesters, from spring 2009 to spring 2011, of E&M being studied. The standard deviation is the standard deviation of the Pearson correlations for those five semesters. The variables with the highest correlations are in bold with a gray background.

| Variable | Avg Correlation | St Dev |
| --- | --- | --- |
| SAT Math | 0.37 | 0.03 |
| ACT Math | 0.44 | 0.02 |
| ACT Science | 0.38 | 0.08 |
| FCI Pre-Test | 0.44 | 0.08 |
| FCI Post-Test | 0.56 | 0.06 |
| CSEM Pre-Test | 0.40 | 0.05 |
| **Calculus 1 Grade** | **0.62** | **0.04** |
| **Calculus 2 Grade** | **0.60** | **0.04** |
| **Mechanics Grade** | **0.65** | **0.08** |
| **Calc 1 + Calc 2 + Mechanics Grade** | **0.72** | **0.04** |

TABLE III. Different physics aptitude groups as determined by students' grade point averages in Calculus 1, Calculus 2, and Mechanics.

| Grade Point Avg | Physics Aptitude Group |
| --- | --- |
| 4.00 - 3.25 | High |
| 3.24 - 2.50 | Medium-High |
| 2.49 – 1.75 | Medium-Low |
| 1.74 - 1.00 | Low |

## IV. RESULTS

Fig. 1 shows plots of total exam score (the average, weighted by percentage of overall course grade, of the scores on the midterm exams plus the final exam) versus homework completion score for different aptitude students who took E&M in the fall 2010 and spring 2011 semesters. The data from these two semesters is typical of the data for the other fall and spring semesters that were analyzed in this study. Included with the plots are line fits for the data. As mentioned in Section II, the fall semesters of E&M have 2-3 times greater enrollment than the spring semesters, so there is much more data for the fall 2010 semester compared to the spring 2011 semester. Remarkably, medium-low and low aptitude students had a negative benefit from doing homework for both semesters shown in Fig. 1, indicating that, on average, the more homework that medium-low and low aptitude students did, the worse they performed on exams. This negative benefit effect for medium-low and low aptitude students also occurred in the other three semesters not shown in Fig. 1. Table IV summarizes the line fit data shown in Fig. 1 for all semesters included in this study and also gives the Pearson correlation and the *p*-value for each of the data sets. The *p*-values from Table IV show that the negative correlations between homework and exam scores for medium-low and low aptitude students are statistically significant ($p < 0.05$) for 4 out of the 5 semesters. Fig. 2 illustrates the difference in benefits that students of various aptitudes got from completing homework by plotting the averages of the slopes of the line fits for the five semesters being studied. Fig. 2 shows that, on average, medium-low and low aptitude students had a negative benefit from completing homework of more than one standard deviation. Finally, Fig. 3 shows the benefit from completing homework for the various aptitude groups on the conceptual multiple-choice questions and on the exam workout problems. While there are differences between the two plots, the relationship between the average slopes and student aptitude is arguably similar for the conceptual multiple-choice questions and the exam workout problems.

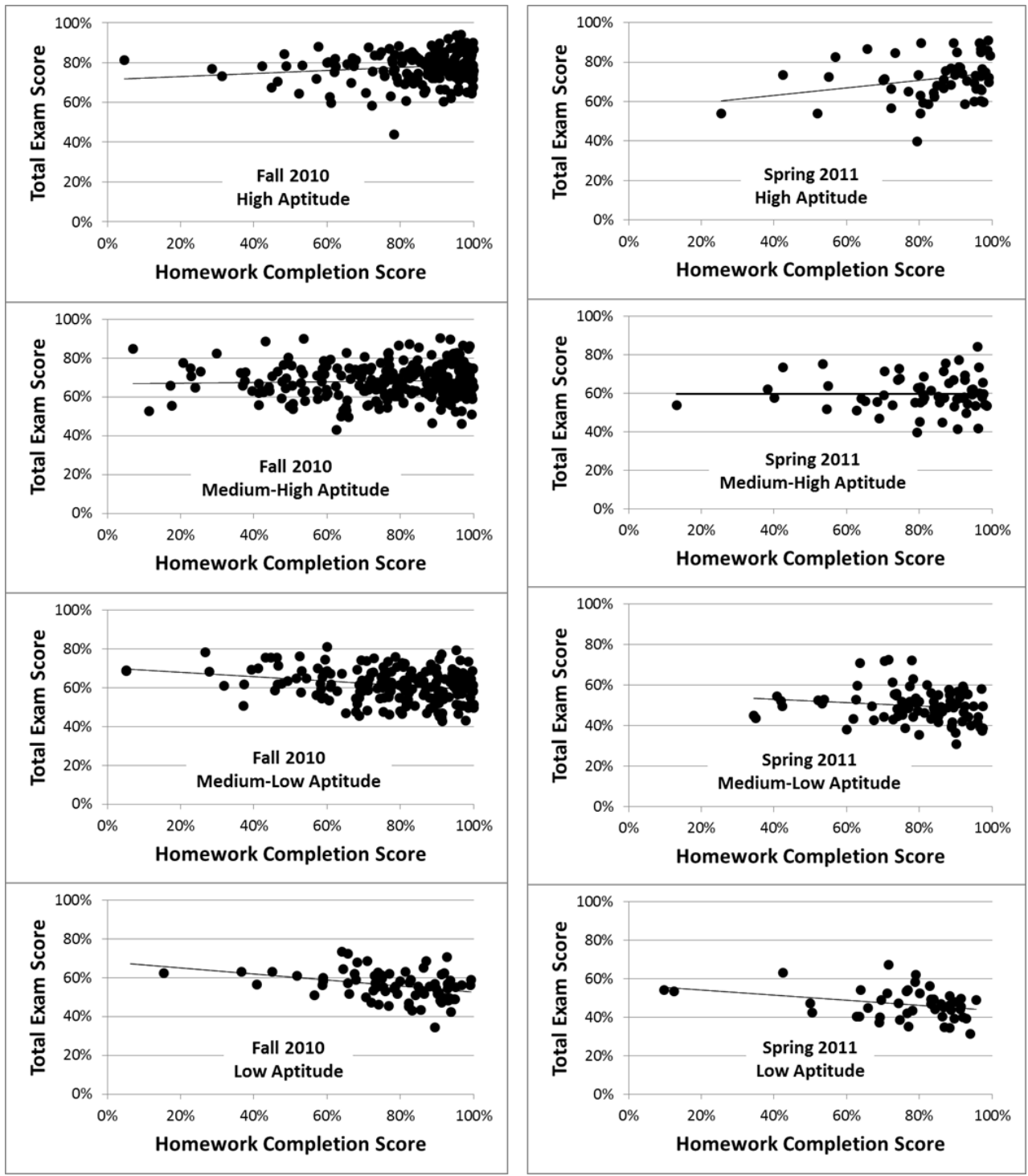

FIG. 1. Plots of the weighted-average score on all exams versus homework completion score for students in different aptitude groups who took the USAFA E&M course in fall 2010 (left plots) and spring 2011 (right plots). Line fits are included with the data.

TABLE IV. Summary of the slopes of the line fits of data plots like those shown in Fig. 1, the Pearson correlations between total exam scores and homework scores, and the *p*-values for those correlations. Statistically significant *p*-values ($p < 0.05$) are in bold and italics.

| Semester | High Aptitude | | | Medium-High Aptitude | | | Medium-Low Aptitude | | | Low Aptitude | | |
|---|---|---|---|---|---|---|---|---|---|---|---|---|
| | Slope | Correlation | *p*-value | Slope | Correlation | *p*-value | Slope | Correlation | *p*-value | Slope | Correlation | *p*-value |
| Spr 09 | 0.30 | 0.295 | 0.153 | -0.03 | -0.075 | 0.692 | -0.12 | -0.390 | ***< 0.001*** | -0.03 | -0.100 | 0.468 |
| Fall 09 | 0.18 | 0.174 | ***0.006*** | -0.10 | -0.179 | ***0.004*** | -0.13 | -0.168 | ***0.018*** | -0.13 | -0.317 | ***0.007*** |
| Spr 10 | 0.05 | 0.072 | 0.604 | -0.13 | -0.259 | 0.070 | -0.14 | -0.265 | ***0.007*** | -0.22 | -0.478 | ***< 0.001*** |
| Fall 10 | 0.08 | 0.150 | ***0.038*** | 0.02 | 0.045 | 0.468 | -0.12 | -0.255 | ***< 0.001*** | -0.16 | -0.332 | ***0.004*** |
| Spr 11 | 0.20 | 0.314 | ***0.032*** | 0.00 | -0.001 | 0.994 | -0.08 | -0.148 | 0.174 | -0.13 | -0.314 | ***0.032*** |

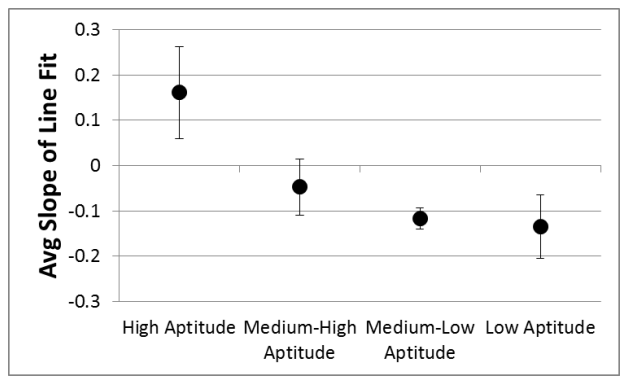

FIG. 2. Average slopes of line fit data as a function of student aptitude for the five semesters of E&M being studied. Error bars are the standard deviations of the slopes for the different semesters.

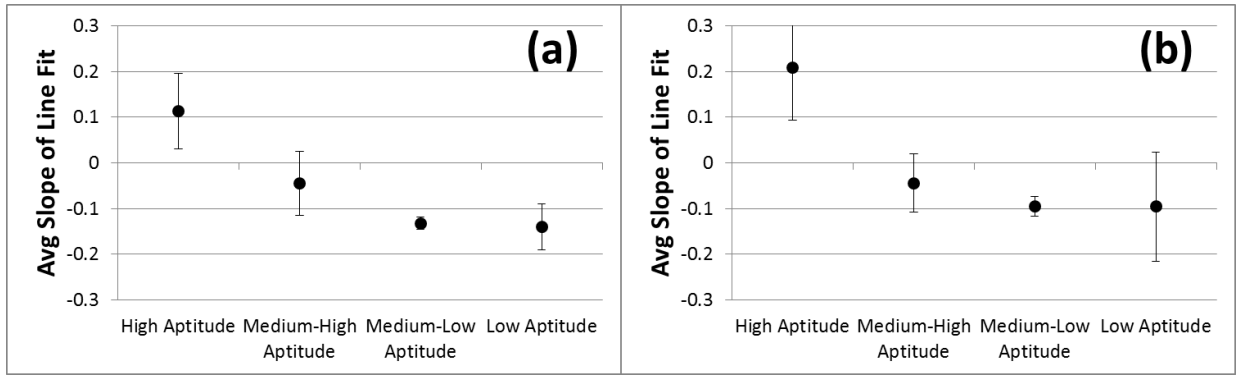

FIG. 3. Average slopes of line fit data as a function of student aptitude for the five semesters of E&M being studied when total exam scores are separated into (a) total score on the conceptual multiple-choice questions on exams, and (b) total score on the workout problems on exams. Error bars are the standard deviations of the slopes for the different semesters.

## V. DISCUSSION

As stated in Section I, homework is typically viewed as an important tool for learning introductory physics. However, the data presented in Section IV indicate that few students benefitted from completing homework and that medium-low and low aptitude students did worse on exams when they completed more homework. In this section, we consider three possible explanations for these data: (1) traditional end-of-chapter homework problems place an excessive cognitive load on lower aptitude students; (2) when lower aptitude students learn magnetism concepts, they have difficulty differentiating between those concepts and electricity concepts, resulting in a negative learning effect for both sets of concepts; (3) a large portion of the lower aptitude students copied their homework from others and gained very little benefit on exams as a result.

### A. Cognitive load

Cognitive load theory [20] states that learners can simultaneously process only a limited number of ideas in their working memory. Additional processing power is available from long-term memory, provided that students have that knowledge encoded in long-term memory. Knowledge thus stored in long-term memory is known as a schema.

High aptitude students appear to possess a large number of useful schemas, both from their previous courses and from an ability to better incorporate early course material into new schemas. As homework problems become more complicated, high aptitude students who call upon these schemas can supplement their limited working memory, enabling them to process the homework and use it to learn new material. On the other hand, low aptitude students likely possess only a limited number of schemas from previous courses, a significant number of which may even be incorrect. Thus, they can only supplement their working memory with a very small number of correct schemas. Because of this, low aptitude students quickly experience cognitive overload. Students may deal with this cognitive overload by creating new, incorrect schemas each time they do their homework. Inevitably, the faulty schema-building that occurs in doing homework problems hinders their success on exams. Low aptitude students who do not do homework will not develop these incorrect schemas and are therefore able to perform better on exams. This will have a cumulative effect, because a proper understanding of physics requires students to build a coherent framework of concepts during the semester. A student who has separate, often incorrect, schemas for all of the different homework problems will have increasing difficulty recognizing the conceptual framework that is being built over the semester. Therefore, though they might be able to memorize and/or pattern-match enough to do adequately on an exam that tests a limited number of concepts, they will fail to achieve the big picture understanding that is necessary to do well on a more comprehensive test, such as the final exam.

To test this theory, we analyzed how homework benefitted different aptitude students on each of the three midterms and compared that to the benefits that those same students got from completing homework when they took the final exam. Specifically, we looked at the benefit on the first midterm exam from completing the homework for lessons 1-10, and similarly, the benefit on the second midterm exam from completing the homework for lessons 11-19, and finally, the benefit on the third midterm exam from completing the homework for lessons 20-30. We then compared each of those to the benefit on the final exam from completing the homework for the entire semester. We did not consider the spring 2009 E&M course in this analysis because there were two, rather than three, midterm exams that semester. Fig. 4 shows the results of this analysis. There is very little difference in benefit between students in different aptitude groups on the first two midterms. On the third midterm, students in the high aptitude group separated themselves from the other students, while students in the low aptitude group had a negative benefit from completing homework. On the final exam, students in both the medium-low and low aptitude groups had a statistically significant negative benefit from completing homework.

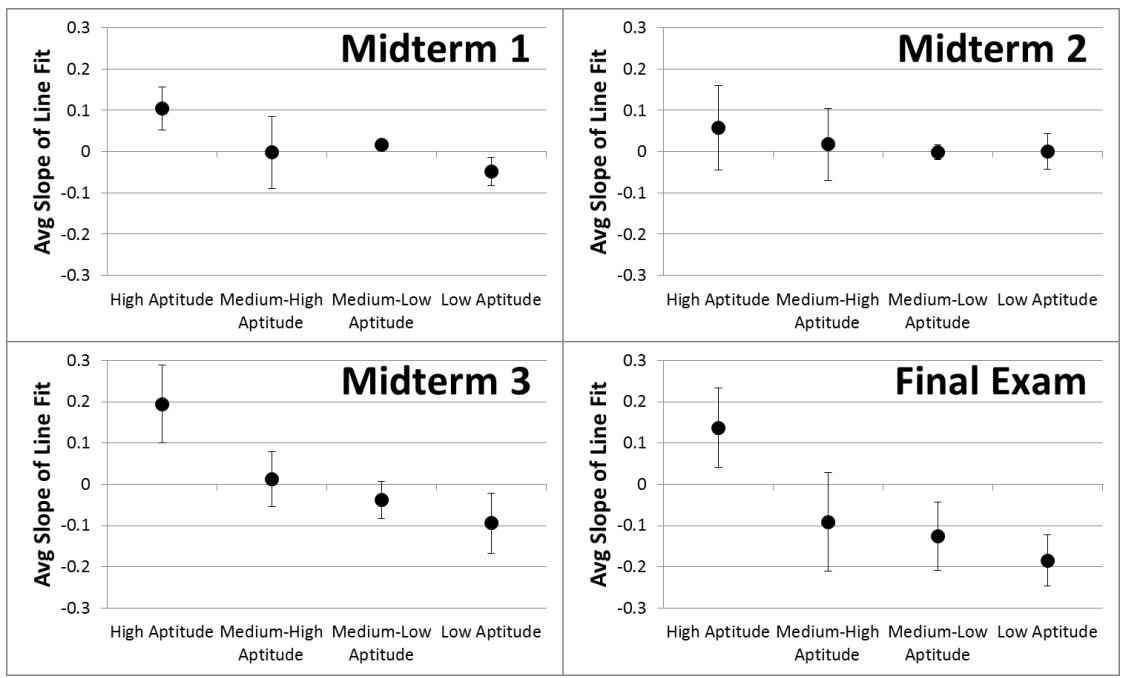

FIG. 4. Average slopes of line fit data as a function of student aptitude for the 4 semesters of E&M from fall 2009 to spring 2011. The line fits, similar to those in Fig. 1, are for the scores on the midterms that happen at the end of each class block (see Table I) versus the homework completion score for that class block. The final exam data is based on plots of final exam scores versus the homework completion scores for the entire semester.

These data suggest that the effects of doing homework are indeed cumulative. High aptitude students used homework as a tool with which to build on their existing schemas to understand and integrate new material. As the course progressed, the schemas of the high aptitude students expanded, providing a larger mental base on which homework can build, making homework a useful tool. In this way, homework was an opportunity for high aptitude students to learn new material while increasing their mastery of old material. For lower aptitude students who had insufficient or incorrect schemas, homework imposed an excessive cognitive load which resulted in confusion rather than learning. It is interesting that the data for the third midterm is different than the data for the other two midterms. This will be discussed further in the next subsection, but it should be noted that there are a greater variety of concepts covered in the third block (see Table I) than in the first two blocks. Also, workout problems on the third midterm often involved concepts from the first two blocks (e.g., finding the cyclotron radius of a charged particle after it has been accelerated through a potential difference). So, it is not surprising that the third midterm may have been difficult for students who had trouble building a conceptual framework.

The fact that homework completion had little effect for medium aptitude students can be explained by viewing the effect of homework on their exam performances as the average of the high aptitude and low aptitude cases. In some situations, medium aptitude students had useful schemas for doing homework problems, and, in those situations, they used homework to build new correct schemas. In other situations, medium aptitude students experienced cognitive overload which resulted in the creation of incorrect schemas. These two effects would tend to counter-balance each other.

### B. Interference between electricity and magnetism concepts

Heckler and Sayre [21] provide another possible explanation for why low aptitude students had a negative benefit from completing homework on the third midterm but not on the first two midterms. The third midterm is the first midterm to cover magnetism concepts, and Heckler and Sayre found that learning about magnetic forces can cause students to develop misunderstandings about electric forces, and that learning about electric forces can cause students to develop misunderstandings about magnetic forces. It would not be surprising if these effects were even more pronounced among lower aptitude students. So, for example, a low aptitude student who spent a lot of time working on homework problems to learn about electric fields could experience great confusion when they try to learn about magnetic fields. Conversely, a low aptitude student who did not do much homework and had only a superficial understanding of electric fields could more easily make the intellectual adjustment necessary for learning about magnetic fields. For medium-low aptitude students, this effect may not become manifest until they are explicitly forced to distinguish between the properties of the two types of fields on the final exam. While there is not enough evidence in Fig. 4 to definitively claim that such an effect is occurring, it is consistent with the data and provides a compelling explanation for why the homework benefit data for the third midterm behave differently than the data for the first two midterms.

### C. Homework copying

Several researchers have studied copying of physics homework [7-8]. Palazzo *et al*. were able to detect copying in Mastering Physics (the same online system used in this study) by examining how long it took for students to answer a question after opening it in the program [7]. Students who submitted a correct answer very quickly (< 3 minutes) after opening a question were identified as copiers because 3 minutes is typically not enough time to read a question, think about it, and submit an answer in Mastering Physics. However, the way homework was done in the USAFA E&M courses was different from what is discussed in the Palazzo *et al.* study because the Mastering Physics problems for the USAFA courses were taken from the textbook (sometimes with the numbers in the problems changed). Therefore, it is possible for students to do considerable work on solving a problem before opening it up in Mastering Physics. Nevertheless, the testing done in that study to detect copying may still be valid for the USAFA E&M courses. To determine if that is the case, we plotted exam scores versus the average time for completion of homework problems in Mastering Physics for medium-low and low aptitude students who took E&M in fall 2010. These data, shown in Fig. 5, do not indicate any relationship between exam scores and the average time it took for students to complete each Mastering Physics problem. This is not very surprising because only two students in the medium-low aptitude group and no students in the low aptitude group would have been identified as consistent copiers using Palazzo *et al.*'s criterion – an average completion time for homework problems that is less than 3 minutes. Additionally, the two students who would have been identified as copiers both scored higher on their exams, 63% and 69%, than the average of the rest of the students in the medium-low aptitude group, 61%. Based on these data, either there were very few homework copiers among the medium-low and low aptitude students in the USAFA E&M course, or the time spent per problem is not a good way to identify homework copiers in this particular sample of students.

An article by Grams discusses a group of students who were doing very well on homework but performed poorly when given similar problems on exams [8]. Additional inquiry revealed that that subset of students achieved their high homework scores by copying solutions from the Internet. We looked for this effect by removing all students with high homework completion rates (> 90%) from the data sets and re-calculating the line fits. If some of the students with high homework completion rates were affecting the data by copying answers from other sources, then we would expect to see noticeably different results when their scores were removed from the analysis. However, Fig. 6 shows that the negative benefit results are essentially unchanged when the data are re-analyzed in this fashion. Consequently, this analysis does not reveal any evidence that homework copying is the reason for lower aptitude students getting negative benefit from completing homework.

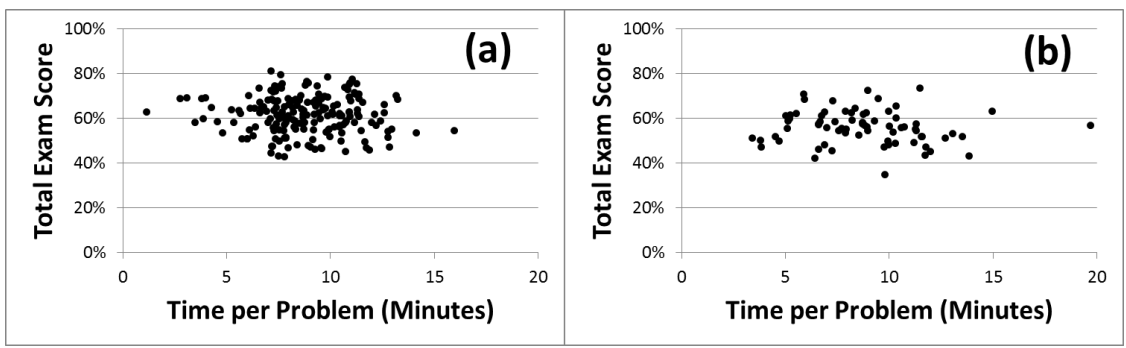

FIG. 5.   Average on E&M exams as a function of time spent per homework problem for (a) medium-low and (b) low aptitude students during the fall 2010 semester.

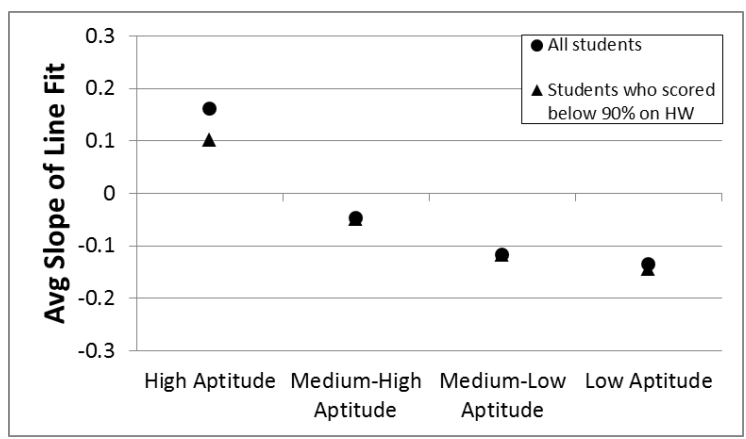

FIG. 6. Average slopes of line fit data as a function of student aptitude for the 5 semesters of E&M from spring 2009 to spring 2011. The circles are the same data shown in Fig. 2, while the triangles are a re-calculation of that data when students who scored above 90% on homework completion are removed from the analysis.

## VI. CONCLUDING REMARKS

We analyzed the benefit that students with different physics aptitudes received from completing homework by plotting their exam scores versus their homework scores for five different semesters of the USAFA introductory E&M course. It was surprising to find that only high aptitude students seemed to derive any measurable benefit from completing homework. Even more surprising, students in the lowest two aptitude groups showed a negative benefit from completing homework. This result is troubling because doing homework is typically considered a key to success in physics courses, so much so that struggling students are often advised to do more homework in order to be better prepared for exams. Our findings indicate that this is ineffective advice and, in fact, that this learning strategy is counterproductive for such students.

We presented two possible explanations for why medium-low and low aptitude students received negative benefit from completing homework. Applying cognitive load theory, we hypothesized that many of the homework problems that were assigned in the E&M course imposed an excessive cognitive load on lower aptitude students. This excessive cognitive load inhibited lower aptitude students' ability to build a conceptual framework and see the big picture of what is going on in the course, resulting in them being poorly prepared for the cumulative final exam and, to a lesser extent, for the third midterm, which covered a greater variety of concepts than the first two midterms. Another possible explanation is based on the fact that it was not until the third midterm and the final exam that noticeable negative benefits from completing homework occurred for lower aptitude students. A previous study found that instruction on magnetic fields can interfere with student understanding of electric fields, and vice-versa. Therefore, it might be the case that the effect we are seeing is due, at least in part, to students' struggle to reconcile the concept of magnetic fields with their previously-established ideas about electric fields. The more homework that students do, the more that each set of concepts come into conflict with each other. Finally, we found no evidence that homework copying had a measurable effect on exam performance for lower aptitude students. At present, cognitive load theory and interference effects between electricity and magnetism concepts provide the best explanations that we have found for why lower aptitude students received a negative benefit from doing homework.

**Notes**

**Acknowledgements**

We would like to thank Dr. Lauren Scharff in the Scholarship of Teaching and Learning Center at the United States Air Force Academy for her support and guidance in this research.